
\input phyzzx
\input tables
%

\def\Ffour{F_{\rm 4}}
\def\Esix{E_{\rm 6}}

\def\hc{{\rm h.c}}
\def\A{{\rm A}}

\def\T{{\rm T}}
\def\one{{\bf 1}}

\def\onezero{{\bf 10}}
\def\onesix{{\bf 16}}
\def\twoseven{{\bf 27}}
\def\threefive{{\bf 351}}
\def\sgn{{\rm sgn}}

\def\Ln{{\rm Ln}}
\def\HurokuMaxLittle{C}
\def\TableESixFour{Table 1}

\def\bar#1{\overline{#1}}
\def\int{\intop\nolimits}

\def\mymatrix#1#2#3#4{\left( \matrix{#1  &  #2  \cr
                                      #3  &  #4  \cr} \right) }
\def\sanmatrix#1#2#3#4#5#6#7#8#9{\left( \matrix{#1  &  #2  & #3 \cr
                #4  &  #5  & #6 \cr #7  &  #8 & #9 \cr} \right) }
\def\myvector#1#2{\pmatrix{#1 \cr #2  \cr}}
\def\sanvector#1#2#3{\pmatrix{#1 \cr #2  \cr #3 \cr}}
\def\gam#1{\,{}^{#1}\kern-1pt\Gamma }
\def\sgm#1{\,{}^{#1}\!\sigma }
\def\barsgm#1{\,{}^{#1}\!\bar\sigma }
\def\Sgm#1{\,{}^{#1}\!\Sigma }
\def\CC#1{\,{}^{#1}\!C }

\def\uni#1{{\bf 1}_{#1}}

\def\ee{\eqno\eq }

\catcode`@=11
\newtoks\KUNS
\newtoks\HETH
\newtoks\monthyear
\Pubnum={KUNS~\the\KUNS\cr HE(TH)~\the\HETH}
\monthyear={\monthname,\ \number\year}
\KUNS={0000}
\HETH={00/00}
\def\p@bblock{\begingroup \tabskip=\hsize minus \hsize
   \baselineskip=1.5\ht\strutbox \topspace-2\baselineskip
   \halign to\hsize{\strut ##\hfil\tabskip=0pt\crcr
   \the\Pubnum\cr \the\monthyear\cr hep-ph/9402357\cr}\endgroup}
\def\bftitlestyle#1{\par\begingroup \titleparagraphs
     \iftwelv@\fourteenpoint\else\twelvepoint\fi
   \noindent {\bf #1}\par\endgroup }
\def\title#1{\vskip\frontpageskip \bftitlestyle{#1} \vskip\headskip }
%
%
\def\Kyoto{\address{Department of Physics,~Kyoto University \break
                            Kyoto~606,~JAPAN}}

%
%
\paperfootline={\hss\iffrontpage\else\ifp@genum%
                \tenrm --\thinspace\folio\thinspace --\hss\fi\fi}
\footline=\paperfootline
%
%

%
\def\journal#1&#2(#3){\begingroup \let\journal=\dummyj@urnal
    \unskip, \sl #1\unskip~\bf\ignorespaces #2\rm
    (\afterassignment\j@ur \count255=#3) \endgroup\ignorespaces }
\def\andjournal#1&#2(#3){\begingroup \let\journal=\dummyj@urnal
    \sl #1\unskip~\bf\ignorespaces #2\rm
    (\afterassignment\j@ur \count255=#3) \endgroup\ignorespaces }
\def\andvol&#1(#2){\begingroup \let\journal=\dummyj@urnal
    \bf\ignorespaces #1\rm
    (\afterassignment\j@ur \count255=#2) \endgroup\ignorespaces }
\def\NP{Nucl.~Phys. }
\def\PR{Phys.~Rev. }

\def\PTP{Prog.~Theor.~Phys. }

\def\MPL{Mod.~Phys.~Lett.}

%
\def\acknowledge{\par\penalty-100\medskip \spacecheck\sectionminspace
   \line{\hfil ACKNOWLEDGEMENTS\hfil}\nobreak\vskip\headskip }
\catcode`@=12

\REF\RMichel{L.~Michel  \journal Rev.~Mod.~Phys. &52 (80) 617;
 Marseille Colloq. (1979) 157.
}

\REF\RGUT{For a review see
 P.~Langacker \journal Phys.~Rep. &72 (81) 187,
and references therein.}

\REF\RTC{
    A.~Farhi and L.~Susskind \journal Phys.~Rep. &74C (81) 277.
    }

\REF\RNJL{Y.~Nambu and G.Jona-Lasinio \journal Phys.~Rev. &112 (61)
345.}

\REF\RTumbling{
    S.~Raby, S.~Dimopoulos and L.~Susskind \journal Nucl.~Phys. &B169
(80) 373.
    }

\REF\RMTY{
    V.A.~Miransky, M.~Tanabashi and K.~Yamawaki
    \journal Phys.~Lett &B221 (89) 177;
    \andjournal \MPL &A4 (89) 1043.
    }

\REF\RBHL{
    W.~Bardeen, C.~Hill and M.~Lindner \journal \PR &D41 (90) 1647.
    }

\REF\RETC{
    S.~Dimopoulos and L.~Susskind \journal Nucl.~Phys. &B155 (79) 237.
    }


\REF\RESix{
    R.~Barbieri and D.V.~Nanopoulos \journal Phys.~Lett. &95B (80) 43;
\nextline
    R.~Barbieri, D.V.~Nanopoulos A.~Masiero
 \journal Phys.~Lett. &104B (81) 194.
    }

\REF\RESixTW{
    P.~Ramond, talk given at Sanibel Symposium (Florida,1979), CALT-68-709.
    }

\REF\RSlansky{
    R.~Slansky \journal Phys.~Rep &79 (81) 1.
    }

\REF\RHarvey{
    A.~Harvey \journal Nucl.~Phys. &B163 (80) 254.
    }

\REF\RGeorgi{
    H.~Georgi \journal Nucl.~Phys. &B159 (79) 126.
    }

\REF\refKT{
    T.~Kugo and P.~Townsend \journal  \NP &B221 (83) 357.
    }

\REF\refIKK{
    K.~Itoh, T.~Kugo and H.~Kunitomo \journal  \PTP &75 (86) 386.
    }

\KUNS={1251}     
\HETH={94/3}     
\monthyear={February, 1994}

\titlepage

\title{ Dynamical Symmetry Breaking in an $E_6$ GUT Model }

\author{Taichiro Kugo and Joe Sato}

\Kyoto  

\abstract{Dynamical symmetry breaking is studied in an $E_6$
GUT model of a single generation of fermions with strong 4-fermi
interactions. The effective potential is analyzed
analytically by the help of Michel's conjecture\refmark{\RMichel}
and the result is confirmed numerically.
We find that the $E_6$ symmetry is spontaneously
broken either to $F_4$ or to
$Sp(8)\ \hbox{or}\ G_{\rm 2}\ \hbox{or}\ SU(3)$, depending on which
of the 4-fermi coupling constants $G_{27}$ and $G_{351}$ in the
\twoseven /\threefive\ channels is stronger. The possibilities
for obtaining other type of breaking patterns are also discussed.
}

\endpage          

\chapter{Introduction}

Despite remarkable successes of the standard model based on
$SU(3)\times SU(2)\times U(1)$, many physicists believe that there
exists a
more fundamental theory beyond it. The strongest evidences for
such theories, usually called grand unified theories
(GUT's)\refmark{\RGUT}, are
the facts that the quark/lepton charges are quantized and that anomaly
is cancelled miraculously between quarks and leptons.
In the usual scenario of GUT's, however, the spontaneous symmetry
breaking required there is discussed by introducing some elementary
Higgs fields just in the same manner as in the standard model.
Then quite a large arbitrariness appears in, \eg, which representations
and how many we introduce as the Higgs fields. Moreover
this introduces too many arbitrary parameters, even more than in
the standard model, in the Higgs Yukawa- and self-couplings.

Dynamical symmetry breaking scenario\refmark{\RTC - \RETC}
 is very attractive in this
respect. There one supposes that there exist only matter fermion
fields belonging to some representation of a gauge group $G$
and the gauge fields of that group. Then the Lagrangian is uniquely
determined by the gauge symmetry alone when we require
the renormalizability (and if the fermions are all chiral).
The usual Higgs fields are supplied as bound states of the fundamental
fermions which are formed by the gauge interaction dynamics itself.
So, which types of Higgs fields appear is determined dynamically
and all the parameters concerning the Higgs fields,
which are arbitrary in the usual scenario, becomes in principle
calculable.

Even when it is difficult to solve fully the dynamics,
the dynamical symmetry breaking scenario can give several constraints
on the possible models for the GUT's, \eg, on possible GUT groups
and/or matter contents. For instance, as was emphasized by
Barbieri and Nanopoulos\refmark{\RESix} and Ramond\refmark{\RESixTW},
$E_6$ is uniquely selected
among many GUT groups if we require i) every generation of
quarks/lepton fields belongs to a single irreducible representation of
the group, ii) the theory is automatically anomaly free, and iii)
all the (phenomenological) Higgs fields necessitated for causing the
symmetry breakings down to $SU(3)_{\rm c}\times U(1)_{\rm em}$
fall in the representations which can be supplied by the fermion
bilinears.  Therefore it is very important to investigate GUT's
from the viewpoint whether they are compatible
or not with dynamical symmetry breaking.

In this paper we study dynamical symmetry breaking in an $E_6$ GUT
model.  The reason why we adopt $E_6$ is its unique property
stated above. In particular the third point implies the possibility
that all the Higgs fields necessary for the symmetry breakings can
be formed dynamically as fermion bound states.  The unified gauge
coupling constant of $E_6$ suggested by the present
experimental data, however,
seems not large enough to break the $E_6$ symmetry
itself, and so we expect that some strong gauge interaction yet
other than the $E_6$ one exists and gives a primary driving force for
the $E_6$ symmetry breaking.
But we still have no definite idea about that gauge interaction beyond
$E_6$.  So we assume in this paper that the strong gauge interaction
is effectively treated
as a Nambu--Jona-Lasinio type 4-fermi interaction[\RNJL].
We include all possible $E_6$-invariant 4-fermi interaction terms that
can contribute to the formation of scalar Higgs fields. We, however,
restrict ourselves to the model of a single generation of
quarks/leptons, with the hope that the Higgs fields are all supplied
as bound states of mainly a single generation of fermions.
Following the usual procedure we introduce Higgs fields
as auxiliary fields. We analyze the effective potential to find the
the patterns of dynamical symmetry breaking realized in this model,
and see whether the desirable symmetry breaking patterns emerge or not.

This paper is organized as follows. In Sect.2 we present the model
which we study in this paper and give the effective potential of
the auxiliary Higgs fields. The analysis of the effective potential
is performed analytically in Sect.3. In a special case in which the
4-fermi interaction is present only in $E_6$ {\bf 27} channel,
a complete analysis is possible and is given there. Otherwise,
however, such a direct analysis becomes almost impossible and we
perform a simplified analysis {\it assuming} that Michel's
conjecture concerning the potential minimum holds.  The symmetry
breaking patterns found this way are actually confirmed to be correct
by numerical analysis performed in Sect.4.
Sect.5 is devoted to the summary and conclusion. Three appendices are
supplemented; by using the spinor representation of $SO(10)$ presented
in Appendix~A,  representations \twoseven\ and \threefive\ of $E_6$
and an invariant tensor with three \twoseven\ indices are explicitly
constructed in Appendix~B; definition of maximal little groups
which appears in Michel's conjecture is presented in Appendix~C.

\chapter{The Model}

As explained in the Introduction, we
consider Nambu--Jona-Lasino type model with a single
generation of left-handed fermions,
$\psi =(\psi _A)\ (A=1,\cdots ,27)$, belonging to {\bf 27}
representation of $\Esix$.
The Langrangian is given in the most generic form as follows:
$$
\eqalign{
{\cal L}\ &=\ {\cal L_{\rm 0}}\ +\ {\cal L_{\rm int}}\cr
{\cal L_{\rm 0}}\
  &=\ \bar \psi i \gamma ^\mu(\partial _\mu - i g A_\mu)\psi
  \ +\ \tr (F^{\mu \nu} F_{\mu \nu}) \cr
{\cal L_{\rm int}}\ &=\ +\ G_{27}|\psi ^T {\rm C} \psi|^2{}_{\bf 27}
 \ +\ G_{351_{\rm S}}|\psi ^T {\rm C} \psi|^2{}_{{\bf 351}_{\rm S}}
 \ +\ G_{351_{\rm A}}|\psi ^T {\rm C} \psi|^2{}_{{\bf 351}_{\rm A}} \ .
}\eqn\Lag
$$
In this expression C denotes the charge conjugation matrix of Lorentz
spinor so that $\psi ^{\rm T}{\rm C}\psi $ is Lorentz scalar,
$G_i$'s denote coupling constants,
subscripts such as {\bf 27} mean the projection into the denoted
irreducible component of $\Esix$ constructed with the fermion
bilinear: ${\bf 27}\times {\bf 27} = \overline{\bf 27}
+ \overline{\bf 351}_{\rm S} + \overline{\bf 351}_{\rm A}$.
The absolute squares are understood to denote $\Esix$-invariant
contractions between those irreducible components and their
complex conjugates. For the present case of single generation fermions
the fermion bilinear $\psi_A^\T C \psi_B$ is symmetric with respect to
the indices $A$ and $B$, and so the last anti-symmetric
component $(\psi_A^\T C \psi_B)_{{\bf 351}_\A}$ vanishes identically.
Henceforth \threefive\ without subscript always denotes
${\bf 351}_{\rm S}$.

Now we introduce auxiliary fields
$(H^\dagger_{{\bf 27}/{\bf 351}})_{AB}$ standing for
$-(\psi_A^\T C \psi_B)_{{\bf 27}/{\bf 351}}$
and rewrite the interaction part ${\cal L_{\rm int}}$ into
$$
\eqalign{
{\cal L_{\rm int}}\
&=\ -\left\{(\psi_A^\T {\rm C} \psi_B)_{{\bf 27}}(H_{{\bf 27}})^{AB}
+\hc \right\}
 \ -\ M_{27}^2 \tr (H_{{\bf 27}}^\dagger H_{{\bf 27}}) \cr
 &\ -\left\{(\psi_A^\T {\rm C} \psi_B)_{{\bf 351}}(H_{{\bf 351}})^{AB}
+\hc\right\}
             \ -\ M_{351}^2 \tr (H_{{\bf 351}}^\dagger H_{{\bf 351}})
\ ,\cr
             &M_{27/351}^2 \equiv {1\over G_{27/351}}
\ . }
\eqn\Lint
$$
We evaluate only the fermion one-loop diagram for our effective
potential. That is formally the leading term in $1/N_g$ expansion
if we introduce $N_g$ copies of our single generation of fermions.
We neglect the $E_6$ gauge interaction since it is expected to
be weak. Then the 1-loop effective potential of $H$ is given by
$$
\eqalign{
\phi(H)& = \phi _0(H)+\phi _1(H)\cr
&\phi _0(H) =  M_{27}^2 \tr (H_{{\bf 27}}^\dagger H_{{\bf 27}})\ +
 \ M_{351}^2 \tr (H_{{\bf 351}}^\dagger H_{{\bf 351}}) \cr
&\phi _1(H) =
-4\int^\Lambda \! {d^4p \over i(2\pi)^4} \ln \det (M^\dagger M-p^2)
\cr
&\qquad M=2(H_\twoseven+H_\threefive)\equiv 2H \ .
}
\eqn\EffPot
$$
Here $\int^\Lambda \!d^4p$ denotes that the integral over $p$ is
defined with
an ultraviolet cutoff $\vert p_{\rm E}\vert \leq \Lambda $
after making the Wick rotation to Euclidean momentum $p \rightarrow
p_{\rm E}$.
If the coupling constants are large enough this potential has a minimum
away from the symmetric point $H=0$ and the $E_6$ symmetry is
dynamically broken.
We can determine the direction of the symmetry breaking
by searching a minimum of this potential.

\chapter{Analysis of the Effective Potential}

\section{Case of {\bf 27} interaction only}

We first consider the simplest case in which only the \twoseven\ part
of the 4-fermi interaction is present; namely,
$$
G_{27} \not = 0, \qquad G_{351} = 0 \ .
\ee
$$
Then clearly the Higgs vacuum expectation value (VEV) can appear only
in the \twoseven\ component:
$$
H_{{\bf 351}} =0, \qquad H=H_{{\bf 27}} \ .
\eqn\DefConO
$$

As explained in Appendix B, \twoseven\ representation of $E_6$ is
decomposed into ${\bf 1} + {\bf 16} + {\bf 10}$ under the $SO(10)$
subgroup and the \twoseven\ Higgs field
$H=H_{{\bf 27}}$ is expressed in a `vector' notation as
$$
V \equiv \pmatrix{H_0 \cr H_\alpha \cr H_M \cr}
\qquad
 \matrix{ \cr (\alpha =1,2,\cdots ,16) \cr (M=1,2,\cdots ,10) \cr}
\eqn\HTwSeParam
$$
where the subscripts 0,
$\alpha$ and $M$ stand for the $SO(10)$ singlet,
{\bf 16} spinor and {\bf 10} vector representations, respectively.
$V$ is embedded into the $27\times 27$ matrix $H=H_{{\bf 27}}$ by the
help of the invariant tensor $\Gamma ^{ABC}$ carrying three \twoseven\
indices, which is explicitly given in Appendix B:
$$
H^{AB} =  \Gamma ^{ABC}V_C \ .
\eqn\HMatrixTwSe
$$
(See Eq.(B.11) for the explicit form of this matrix.)

Any \twoseven\ can be $\Esix$ rotated into the following `standard'
reduced-form:
$$
V_0 =
\pmatrix{ v_0 \cr
          {\bf 0} \cr
          R + i I\cr
          m\cr
          {\bf 0}        }
\eqalign{
 \vbox{
  \hbox{$\}SO(10)$ singlet (real)}
  \hbox{$\}SO(10)$ \onesix}
  \hbox{$\}$the first component of $SO(10)$ \onezero\ (complex)}
  \hbox{$\}$the second component of $SO(10)$ \onezero\ (real)}
  \hbox{$\}$the third to tenth components of $SO(10)$ \onezero}
 }
}
\eqn\HRotated
$$
$$
\hbox{($v_0,R,I,m$ are all real)} \ .
$$
This is seen as follows. First, starting from a generic form
\HTwSeParam\  of $V$, the spinor component $H_\alpha $ can be rotated
away by using the $\Esix$ rotation freedom with the spinor parameter
$\epsilon $ in (B.2). Next we note that the vector component
$H_M$ actually stands for two $SO(10)$ irreducible \onezero\ vectors,
the real and imaginary parts. So, using the $SO(10)$ rotation freedom,
we can make one of the two \onezero\ vectors,
say the imaginary parts, to have only the first component, and finally,
by using the remaining $SO(9)$ rotation freedom,
we can make the real part \onezero\ to have only the first two
components. Alternatively, one can also convince the validity of
the statement as follows:
the problem is whether any \twoseven\ $V$ can be written in the
form $gV_0$ with $g\in E_6$ by using the standard form $V_0$
in \HRotated. Note as for $g$ in this expression that only the right
quotient $E_6/SO(8)$ part is effective
since $V_0$ is invariant under $SO(8)$. So the $g$ part is
parameterized
by $78-28=50$ parameters, and hence
$gV_0$ spans a $50+4=54$ dimensional space. But it is
the same dimensions as the whole \twoseven\ complex vector $V$ does.

This standard form \HRotated\ has four parameters and
implies that there exist four $\Esix$-invariants which can be
constructed by \twoseven\ representation $V$ alone. They can easily be
found and are given as follows:
$$
\eqalignno{
&X \equiv V^\dagger V &\eqname\XInv\cr
&\eqalign{Y \equiv \Gamma^{ABC} V_A V_B V_C
\quad\hbox{and its complex conjugate }\quad Y^*} &\eqname\YInv\cr
&Z \equiv \Gamma^{ABC} V_B V_C (\Gamma^{ADE} V_D V_E)^* \ .
&\eqname\ZInv
}
$$
These invariants $X, Y, Y^*$ and $Z$ are expressed in terms of the
four parameters in \HRotated\ as
$$
\eqalign{
X\ &=\ v_0^2+R^2+m^2+I^2\cr
Y\ &=\ v_0(R^2+m^2-I^2+2iRI), \quad Y^*\ =\ v_0(R^2+m^2-I^2-2iRI)\cr
Z\ &=\ \left\{\left (R^2+m^2+I^2\right )^2-4I^2 m^2\right\}
 +4v_0^2(R^2+m^2+I^2) \ .
}\ee
$$

Since the 1-loop effective potential $\phi(H)$ is invariant under
$\Esix$, $\phi(H)$ can be expressed in terms of the invariants,
$X,\ Y\ (Y^*)$ and $Z$ alone. The effective potential \EffPot\ now
reads
$$
\eqalign{
\phi(H=M/2)&= M_{27}^2 {1\over 4}\tr (M^\dagger M)
 -4\int^\Lambda\! {d^4p \over i(2\pi)^4} \ln \det(M^\dagger M-p^2)\cr
 &= {1\over 4} M_{27}^2 \sum_{i=1}^{27} \lambda_i
 -4 \int^\Lambda\! {d^4p \over i(2\pi)^4} \sum_{i=1}^{27}
 \ln (\lambda_i-p^2) \ ,\cr
}
\eqn\EffPotP
$$
where $\lambda_i$'s are real positive eigenvalues of $M^\dagger M$,
given by the roots of the following equations:
$$
\eqalignno{
&\lambda_i^3-8X\lambda_i^2+16X^2\lambda_i-64YY^*= 0
\quad  {\rm for\ }i=1,2,3
&\eqname\EqEigenOTT\cr
&\lambda_i^3-4X\lambda_i^2+4Z\lambda_i-16YY^*= 0
\qquad  {\rm for\ }i=4,\cdots ,27.
&\eqname\EqEigenFTS \cr
}
$$
These equations depend on only three quantities of the four invariants:
$$
\eqalign{
X\ &=\ v_0^2+R^2+m^2+I^2\cr
YY^*\ &=\ v_0^2\big\{(R^2+m^2+I^2)^2-4I^2m^2\big\}\cr
Z\ &=\ \left\{\left (R^2+m^2+I^2\right )^2-4I^2 m^2\right\}
 +4v_0^2(R^2+m^2+I^2) \ . \cr
}\ee
$$
For convenience, we re-parameterize these three quantities as follows:
$$
\eqalign{
&a\equiv v_0^2\cr
&b\equiv R^2+m^2+I^2\cr
&c\equiv Im
}
\ \ \Longleftrightarrow \ \
\eqalign{
&X=a+b\cr
&YY^*=a(b^2-4c^2)\cr
&Z=(b^2-4c^2)+4ab
}
\eqn\Defabc
$$
The three roots of Eq.\EqEigenOTT\ cannot explicitly be expressed
in terms of $a,b$ and $c$,  but the twenty-four roots of
Eq.\EqEigenFTS\ are given by
$$
\eqalign{
&\lambda_i=4a\ \ \ (i=4\ldots 11)\cr
&\lambda_i=2b+4c\ (i=12\ldots 19)\cr
&\lambda_i=2b-4c\ (i=20\ldots 27) \ . \cr
}
\eqn\EigenFTSP
$$
Then, inserting these explicit expressions for the roots
$\lambda _i\ (i=4,\cdots ,27)$ and using an identity
$$
(\lambda _1+y)(\lambda _2+y)(\lambda _3+y)=y^3+8Xy^2+16X^2y+64YY^*
\ee
$$
following from Eq.\EqEigenOTT\ for the implicit roots
$\lambda _{1,2,3}$, the effective potential \EffPotP\ reduces to
$$
\eqalign{
\phi (H) &=\phi (a,b,c)\cr
 &= 10 M_{27}^2(a+b)
 -{1\over 4\pi ^2}\int^{\Lambda^2}\!ydy\Bigl[
\ln(y^3+8Xy^2+16X^2y+64YY^*)
 \cr
 &\qquad\qquad +8\ln (4a+y)+8\ln (2b+4c+y)
 +8\ln (2b-4c+y) \Bigr] \ . \cr
}
\eqn\EffPotPP
$$

We now look for the stationary point of the effective potential
$\phi (H)=\phi (a,b,c)$.  Taking account that $X$ is independent of
$c$ and $\partial (YY^*)/\partial c=-8ac$, the derivative of
the potential $\phi $
with respect to the parameter $c$ is given by
$$
{\partial\phi \over \partial c}
 = - {1\over 4\pi^2} \left[ -64\cdot 8ac\,f(\Lambda^2)
 + 8\cdot 4\bigl\{g(2b+4c)-g(2b-4c)\bigr\}\right]
\eqn\PotPartialC
$$
with functions $f$ and $g$ defined by
$$
\eqalign{
f(x) & = \int _0^x\!dy{y\over y^3+8Xy^2+16X^2y+64YY^*} \cr
g(x) & = \int _0^{\Lambda ^2}\!dy{y\over y+x}
       = \Lambda ^2 - x \ln {x+\Lambda ^2\over x} \ . \cr
}\ee
$$
Note that $f(x)$ is positive for $x>0$ since
$f(0)=0$ and
$$
f'(x) = {x\over x^3+8Xx^2+16X^2x+64YY^*} > 0
\ee
$$
because $X\geq 0$ and $YY^*\geq 0$ by definition \Defabc.
On the other hand, $g(x)$ is seen to be a monotonously decreasing
function of $x$ since
$$
g'(x)= - \int_0^{\Lambda^2}\!{ydy\over (y+x)^2} < 0 \ .
\ee
$$
Taking account also that $c$ is bounded ($|c|<|b|/2$) by definition
\Defabc, we find that
$$
\sgn ({\partial\phi \over \partial c}) = \sgn (c)\ .
\ee
$$
This shows that $\phi(a,b,c)$ has an absolute minimum at
$$
c=0
\eqn\ConTSO
$$
in the defining region $\vert c\vert <\vert b\vert /2$.

Next consider the derivatives of $\phi(a,b,c)$ with respect to $a$
and $b$; at $c=0$ they are given respectively by
$$
\eqalign{
{\partial\phi \over \partial a}
 =10M_{27}^2 - {1 \over 4\pi^2} \biggl [
&\int _0^{\Lambda ^2}\!ydy{8y^2+32Xy\over y^3+8Xy^2+16X^2y+64YY^*}  \cr
 & \qquad + 64\cdot b^2\,f(\Lambda ^2)  + 8\cdot 4\,g(4a) \biggr ]
}
\eqn\PotPartialA
$$
$$
\eqalign{
{\partial\phi \over \partial b}
 =10M_{27}^2 - {1 \over 4\pi^2} \biggl [
 &\int _0^{\Lambda ^2}\!ydy{8y^2+32Xy\over y^3+8Xy^2+16X^2y+64YY^*}  \cr
 & \qquad + 64\cdot 2ab\,f(\Lambda ^2)  + 8\cdot 4\,g(2b) \biggr ] \ .
}
\eqn\PotPartialB
$$
Stationarity requirement
$\partial \phi /\partial a=0$ and $\partial \phi /\partial b=0$, or
only the difference
$\partial \phi /\partial a-\partial \phi /\partial b=0$,
leads to the following condition:
$$
2b(2a-b)f(\Lambda^2) =g(4a)-g(2b) \ .
\ee
$$
Since $a,b>0$ by definition and $f(\Lambda^2)>0$ as mentioned above,
the sign of the LHS is $\sgn(2a-b)$
while the sign of the RHS is opposite, $-\sgn(2a-b)$,
since $g(x)$ is a monotonously decreasing function.
Therefore $\phi(a,b,c)$ can have a minimum only at
$$
2a=b\ .
\eqn\ConTST
$$

Under the conditions \ConTSO\ and \ConTST, the \twoseven\ vector
\HRotated\ now takes the form:
$$
\eqalign{
V =
\pmatrix{ v \cr
          0 \cr
          \sqrt 2 v\cr
          0
 }
}
\eqalign{\vbox{\hbox{$\}SO(10)$ singlet (real)}
 \hbox{$\}SO(10)$ {\bf16}}
 \hbox{$\}$the first component of $SO(10)$ {\bf 10} (real)}
 \hbox{$\}$the second to tenth components of $SO(10)$ {\bf 10}}
 }}
\eqn\HConcluded
$$
For the VEV of this form, the 27 eigenvalues $\lambda _i$
of the fermion squared mass matrix $M^\dagger M$, determined
by \EqEigenOTT\ and \EqEigenFTS, now become explicit and show an
interesting degeneracy: one $16v^2$ and twenty six $4v^2$'s.
Then the potential \EffPotP\ is given by
$$
\phi (H)
= 30 M_{27}^2v^2
 -{1\over 4\pi ^2}\int^{\Lambda^2}\!ydy\Bigl[ \ln (16v^2+y)+26\,\ln
(4v^2+y)
  \Bigr] \ ,
\ee
$$
the stationarity of which determines the magnitude of the VEV $v$:
$$
30 M_{27}^2 = {1\over 4\pi ^2}\left( 16\,g(16v^2) + 26\cdot 4\,g(4v^2)
\right)\ .
\ee
$$
The critical coupling $G_{27}^{\rm cr} = 1/(M_{27}^{\rm cr})^2$
beyond which non-zero VEV is realized is found by taking $v^2
\rightarrow  0$:
$$
G_{27}^{\rm cr} = {\pi^2 \over \Lambda ^2} \ .
\ee
$$

Note that on this vacuum one fermion has mass $4v$ and all the other
26 fermions have a degenerate mass $2v$. This implies that
the original fermion multiplet \twoseven\ splits into
${\bf 1} + {\bf 26}$. This branching pattern indicates that
the symmetry breaking realized in this case is
$$
\Esix\ \longrightarrow\ \Ffour\ .
\ee
$$
This can also be confirmed by calculating branching pattern of
the $E_6$ gauge boson masses on this vacuum;
${\bf 78} \rightarrow  {\bf 26}({\rm massive}) + {\bf 52}({\rm
massless})$,
where {\bf 52} is massless gauge bosons of the unbroken $F_4$ group.

\section{Case of {\bf 351} interaction only}

Next we consider the case in which only the \threefive\ part
of the 4-fermi interaction is present; namely,
$$
G_{351} \not= 0 , \qquad G_{27} = 0\ ,
\ee
$$
in which case the Higgs VEV appear only in the \threefive\ component:
$$
H_{{\bf 27}} =0, \qquad H=H_{{\bf 351}} \ .
\eqn\DefConO
$$
Contrary to the previous case, there are many $\Esix$ invariants and
it is almost impossible to perform the same kind of analysis as
in the previous subsection.

Now we invoke Michel's conjecture[\RMichel,\RSlansky], which claims
that
the following statement holds
when the potential system contains only a real irreducible
representation
of scalar fields, or a self-conjugate pair of a complex irreducible
representations:
{\it
The group symmetry can break down only to one of
the maximal little groups of the representation considered}.
(See Appendix C for the definition of maximal little groups.)

For illustration, let us apply this conjecture to the previous case;
 there, the fields appearing in the potential are (auxiliary) Higgs
fields of {\bf 27} and its conjugate $\overline {\rm{\bf 27}}$.
Then the maximal little groups are $SO(10)$ and $\Ffour$.
We in fact found the symmetry breaking $\Esix \rightarrow  \Ffour$ in
the
above, so that the conjecture was actually correct.

In the present case of {\bf 351} Higgs fields, the
maximal little groups are $SO(10)$, $\Ffour$,
$Sp(8)$, $G_{\rm 2}$, $SU(3)$ and $SU(2)\otimes SU(4)$.
We assume that the conjecture holds in this case also. Then what we
have to do is to calculate the effective potential
for each possibility of the Higgs VEV's in the maximal little group
directions and to compare the minimum values of the potentials to see
which possibility is actually realized.
\midinsert
\medskip
  \begintable
$H\subset E_6$  \hfill & decomposition under $H$\ \  \cr
$ \Ffour$      \hfill& \ {\bf 27}={\bf1}+{\bf26} \hfill \nr
$ Sp(8)$       \hfill& \ {\bf 27}={\bf 27} \hfill \nr
$ G_{\rm 2}$   \hfill& \ {\bf 27}={\bf 27} \hfill \nr
$ SU(3)$       \hfill& \ {\bf 27}={\bf 27} \hfill \nr
$ SU(2)\otimes SU(4)\ \ \ $\hfill& \ {\bf 27}=({\bf2},{\bf6})+({\bf1},
{\bf15}) \hfill
\endtable
\smallskip
\Table{1}{. \ Decomposition of $E_6$  {\bf 27} under the maximal groups
$H$.}
\medskip\endinsert
Let us first determine the form of the VEV for each case of the
maximal little groups. From \TableESixFour, we see that the fermion
{\bf 27} of $\Esix$ is again {\bf 27}
under $Sp(8)$, $G_{\rm 2}$ and $SU(3)$ but
the latter is a real representation.
Taking also account
that $\Esix$ {\bf 351} is constructed from ${\bf 27} \otimes {\bf 27}$
symmetrically, we see that the
trace part of $H_{{\bf 351}}$ is a singlet under those groups.
Therefore, in a suitable basis, the VEV takes the following form
for the cases of those groups:
$$
H_{{\bf 351}}=v\otimes \one_{27} \qquad
\hbox{($Sp(8)$, $G_{\rm 2}$, $SU(3)$)} \ ,
\eqn\HThFiOSpE
$$
where $\one_{n}$ denotes $n\times n$ identity matrix.

Similarly, in case of $SU(2)\otimes SU(4)$,
a singlet under this group which we get from
a symmetric product of ${\bf 27} \times  {\bf 27}$
comes from the component $({\bf1},{\bf15})\times ({\bf1},{\bf15})$
and hence, in a basis,
$$
H_{{\bf 351}}=\pmatrix{v\otimes \one_{15}&\cr &0\otimes
\one_{12}}\qquad
  \hbox{($SU(2)\otimes SU(4)$)} \ .
\eqn\HThFiOSUF
$$

It is a bit more complicated to get the form of VEV
in cases of $SO(10)$ and $\Ffour$, since, as is seen from
\TableESixFour, we get two singlets from the
symmetric product ${\bf 27} \times  {\bf 27}$
for each case of $SO(10)$ and $\Ffour$.  In view of
${\bf 27} \otimes {\bf 27} |_{\hbox{sym}}= \overline {{\bf 27}}
+\overline {{\bf 351}}$, we see that one singlet is in $H_{{\bf 27}}$
and the other is in $H_{{\bf 351}}$. We can find the form of
$SO(10)$ and $F_4$ singlets in {\bf 27}; from \HMatrixTwSe,
\HRotated\ and \HConcluded, they are given, respectively, by
$$
H_{{\bf 27}}=\pmatrix{
            &0\otimes \one_{17}&\cr
            &&V\otimes \one_{10}}    \qquad \ ( SO(10) ) \ .
\eqn\HTwSeSOTen
$$
$$
H_{{\bf 27}}=\pmatrix{
 &-2V&\cr
 &   &V\otimes \one_{26}
} \qquad \ ( F_4 ) \ .
\eqn\HTwSeFFour
$$
We see from \HTwSeSOTen\ that the $SO(10)$
singlet in $H_{{\bf 27}}$ comes solely from
${\bf 10} \times  {\bf10}$. Therefore the $SO(10)$ singlet in
$H_{{\bf 351}}$ must be the other singlet made from
$\one \times \one $:
$$
H_{{\bf 351}}=\pmatrix{v&\cr& 0\otimes \one_{26}}\qquad
\hbox{($SO(10)$)} \ .
\eqn\HThFiOSOTen
$$
(This can also be seen directly from (B.10) in Appendix B.)
In case of $\Ffour$, the two $F_4$ singlets come from
${\bf 26} \times  {\bf 26}$ and $\one\times \one$, one combination
of which is \HTwSeFFour\ contained in $H_\twoseven$. Since \threefive\
is orthogonal to \twoseven, $\tr H_{{\bf 27}}^\dagger H_{{\bf 351}}=0$,
the $F_4$ singlet in $H_{{\bf 351}}$ should thus have the form:
$$
H_{{\bf 351}}=
\pmatrix{
 &13v&\cr
 &   &v\otimes \one_{26}}\qquad
\hbox{ ( $\Ffour$ )} \ .
\eqn\HThFiOFFour
$$

Now that we have found the form of VEV's, we can calculate the
minimum value of the potential by
 substituting those matrix into \EffPot\ for each case and compare
their minimum values to find the direction of symmetry breaking.
First we define a function
$$
F(v) \equiv M_{351}^2 v^2
- 4 \int^\Lambda\!{d^4p\over i(2\pi)^2}\ln(4v^2-p^2) \ .
\ee
$$
Then the potential value in the directions
$Sp(8)$, $G_{\rm 2}$ and $SU(3)$ is commonly given by
$$
\phi_{Sp}=27F(v)\ .
\eqn\VSp
$$
In the directions $SU(2)\otimes SU(4)$, $SO(10)$ and $F_4$, it is given
respectively by
$$
\eqalignno{
\phi_{SS}&=15F(v)\ ,            &\eqname\VSS \cr
\phi_{SO}&=F(v)\ ,               &\eqname\VSO \cr
\phi_{F4}&=26F(v)+F(13v)\ .      &\eqname\VFF \cr
}$$
When symmetry breaking occurs,
$F(v)$ has a minimum at a certain point $v_0$ and takes a negative
value there.  The potentials \VSp, \VSS\ and \VSO\ take their minima
at the same point $v_0$ and hence we immediately see
for the minimal values
$$
\phi_{Sp}<\phi_{SS}<\phi_{SO}\ .
\ee
$$
The minimum of \VFF\ is realized at a certain point $v_1$, which is
different from the minimum point $v_0$ of $F(v)$, so that
$$
F(v_0)\le F(v_1)\quad \hbox{and}\quad F(v_0)\le F(13v_1) \ ,
\ee
$$
and hence
$$
\phi_{Sp}\le \phi_{F4}\ .
\ee
$$
We thus find that the symmetry breaking in this pure \threefive\
interaction case is
$$
\Esix \longrightarrow Sp(8)\ \hbox{or}\ G_{\rm 2}\ \hbox{or}\ SU(3)\ .
\ee
$$
These three group cases cannot be distinguished in the present
approximation in which only the fermion one-loop vacuum energy is
counted, since the fermions get quite the same masses for those three
breakings. This degeneracy will be lifted if
the vacuum energy due to gauge boson loops are taken into account.

\section{general case}

Finally in this section we study the general case in which
both \twoseven\ and \threefive\ 4-fermi interactions are present.
Strictly speaking,
Michel's conjecture is inapplicable to this general case
since there appear two fields of different representations
{\bf 27} and {\bf 351} in the potential.  Nevertheless
we assume that this conjecture still holds and determine the
symmetry breaking pattern in this case also using
the same analysis method as in the previous subsection.

Candidate groups are the same as the {\bf 351} case:
$SO(10)$, $\Ffour$, $Sp(8)$, $G_{\rm 2}$, $SU(3)$ and
$SU(2)\otimes SU(4)$.
Of these groups $Sp(8)$, $G_{\rm 2}$, $SU(3)$ and $SU(2)\otimes SU(4)$
have their singlet only in {\bf 351} of $\Esix$, and
so the VEV's and potentials are the same as in the previous subsection.
Therefore
$$
\phi_{Sp} < \phi_{SS}
\ee
$$
is always realized.

On the other hand $SO(10)$ and $\Ffour$ have their singlets in both
{\bf 27} and {\bf 351} representations of $\Esix$.
Their singlets in {\bf 351} are contained in the form
\HThFiOSOTen\ and \HThFiOFFour\ and those in in {\bf 27}
are in the form \HTwSeSOTen\ and \HTwSeFFour, for the
$SO(10)$ and $\Ffour$ cases, respectively. Thus the
general forms of VEV's for these group cases are given respectively by
$$
H=\pmatrix{v&&\cr
            &0\otimes \one_{16}&\cr
            &&V\otimes \one_{10}} \qquad ( SO(10) )
\eqn\HGenSOTen
$$
$$
H=\pmatrix{
 &13v-2V&\cr
 &   &(v+V)\otimes \one_{26}} \qquad  ( F_4 ) \ .
\eqn\HGenFour
$$
By using \HGenSOTen,  the potential corresponding to $SO(10)$ breaking
($\equiv \phi_{SO}'$) is
$$
\eqalign{
\phi_{SO}'
=(M_{27}^2-M_{351}^2) V^2 + F(V) +10 F(v)
}
\eqn\VSOO
$$
and by using \HGenFour, that of $\Ffour$ breaking
($\equiv \phi_{F4}'$) is
$$
\eqalign{
\phi_{F4}'= 30(M_{27}^2-M_{351}^2) V^2 + F(13v-2V) +26 F(v+V) \ .
}
\eqn\VFFF
$$

Now let us compare $\phi_{Sp}$,$\phi_{SO}'$ and $\phi_{F4}'$ at their
minimum points.
First of all we clearly see the relation
$$
\phi_{Sp}=\phi_{F4}'<\phi_{SO}'
\ \ \hbox{when}\ \
M_{27}^2 = M_{351}^2 \ .
\eqn\InitCon
$$
In view of this,
we study the potential in two cases of
(a) $M_{27}^2 > M_{351}^2$ and
(b) $M_{27}^2 < M_{351}^2$, separately.

\noindent
(a) $M_{27}^2 > M_{351}^2$

In this case we have from \VSOO
$$
\eqalign{
\phi_{SO}'&=(M_{27}^2-M_{351}^2) V^2 + F(V) +10 F(v)\cr
&> 11F(v_0) \cr
&> 27 F(v_0) = \phi_{Sp} \ , \cr
}\ee
$$
and from \VFFF
$$
\eqalign{
\phi_{F4}'
&= 30(M_{27}^2-M_{351}^2) V^2 + F(13v-2V) +26 F(v+V)\cr
&> F(13v-2V) +26 F(v+V)\cr
&> 27 F(v_0)=\phi_{Sp} \ . \cr
}\ee
$$
Hence we conclude that the symmetry breaking pattern in this case is
given by
$$
\Esix\ \ \longrightarrow\ \
Sp(8)\ \hbox{or}\ G_{\rm 2}\ \hbox{or}\ SU(3)
\ee
$$

\noindent
(b) $M_{27}^2 < M_{351}^2$

Taking account that $\phi_{Sp}$ does not depend on $M_{27}^2$,
we first study the derivative of $\phi_{F4}'$ with respect to
$M_{27}^2$, with $M_{351}^2$ kept fixed.
The arguments $V$ and $v$ of $\phi_{F4}'$ are set equal to the values
realizing the stationary point of $\phi_{F4}'$ and so
they depend on $M_{27}^2$.
$$
\eqalign{
{\partial \phi_{F4}'({\rm stationary\ point})
\over \partial M_{27}^2}
&=30V^2
+{\partial V \over \partial M_{27}^2}
{\partial \phi_{F4}' \over \partial V}
+{\partial v \over \partial M_{27}^2}
{\partial \phi_{F4}' \over \partial v} \cr
&=30V^2 \geq  0 \ . \cr
}\ee
$$
This implies that the minimum value of $\phi_{F4}'$
is monotonously increasing
as a function of $M_{27}^2$, and hence together with \InitCon\ that
$$
\phi_{F4}'<\phi_{Sp}
\eqn\eqINEQUALITYI
$$
in this region $M_{27}^2 < M_{351}^2$.

Next we compare $\phi_{F4}'$ and $\phi_{SO}'$. In the limiting region
$$
M_{351}^2 \gg  M_{27}^2 \rightarrow 0
\qquad \hbox{namely}
\qquad G_{351} \ll  G_{27} \rightarrow \infty \ ,
\ee
$$
the system is the same as that where there is only {\bf 27} 4-fermi
interaction and there, as we know, the $F_4$ vacuum is the lowest one:
$$
\phi_{F4}'< \phi_{SO}' \ .
\eqn\eqINEQUALITYII
$$
On the other hand the relation \InitCon\ implies
that the same inequality holds even in the region
$M_{27}^2 \sim M_{351}^2$.
This strongly suggests that the inequality \eqINEQUALITYII\
holds for the whole region $M_{27}^2 < M_{351}^2$.
We assume this holds.
Then, together with \eqINEQUALITYI, we find that the symmetry
breaking pattern in this coupling region is
$$
\Esix\ \ \longrightarrow\ \ \Ffour \ .
\ee
$$

The discussion in this subsection is very incomplete by two reasons.
Firstly, this general case is outside the scope of Michel's conjecture.
Secondly, the Eq.\eqINEQUALITYII\ was not proved for the whole region
of $M_{27}^2 < M_{351}^2$. Nevertheless it suggests a simple
symmetry breaking pattern; it is either
$\Esix\ \ \rightarrow\ \ \Ffour$  or
$\Esix\ \ \rightarrow\ \ Sp(8)\ \hbox{or}\ G_{\rm 2}\
\hbox{or}\ SU(3)$ depending on whether the \twoseven\ interaction
$G_{27}$ is larger or smaller than the \threefive\ interaction
$G_{351}$, respectively.

\chapter{Numerical Analysis}

In order to confirm the symmetry breaking pattern suggested by
the analysis in the previous section, we numerically search
the minimum of the potential \EffPot\   and calculate
a fermion mass spectrum and gauge boson mass spectrum at that point.

\section{Algorithm}

We present in this subsection an algorithm for searching the stationary
point $H_{\rm st}$: $\partial \phi /\partial H^\dagger|_{H_{\rm
st}}=0$.
The idea is essentially to apply the Newton method to the derivative
$\partial \phi (H)/\partial H^\dagger$ since we want a zero point of
this function.

First of all we note:
\item{1)} $\phi (H)$ is a function of $378\times 2$ variables as $H$ is
a $ 27\times 27$ symmetric and complex matrix.
\item{2)}
$\partial \phi (H)/\partial H^\dagger\equiv V(H)$
is a gradient of $\phi (H)$ in the 756 dimension space,
which can be written down in a closed matrix form:
$$
\eqalign{
V(H)&\equiv {\partial \phi \over \partial H^\dagger }(H) \cr
 &= (M_{27}^2-M_{351}^2)H_\twoseven + M_{351}^2H  \cr
 & \quad -{1\over \pi ^2}H\left[\Lambda ^2-4H^\dagger
H\left(\Ln(4H^\dagger H+\Lambda ^2) -\Ln(4H^\dagger H) \right)\right]
\ . \cr }
\ee
$$
\item{3)} On the contrary we have no such a simple analytic expression
for the second derivative of $\phi(H)$.

\noindent
We now outline how the iteration method goes for searching the minimum.
(We assume in the following for simplicity that
$\phi (H)$ is {\it concave} in the considered region.)

\item{i)}
 We take randomly a starting $H \equiv H_0$, and
 calculate the gradient
$V(H_0)={\partial \phi \over \partial H^\dagger}(H_0)$ there.
\item{ii)}
 To find the next point which is nearer to the stationary point,
 we consider the potential function $\phi (H)$ in a cross section in
 the gradient direction; namely, we consider the following function
 of one real parameter $t$:
$$
f(t)\equiv \phi(H_0+Vt)\ .
\ee
$$
If we find a zero of the first derivative function
$$
g(t)\equiv {d f(t)\over dt}
= \tr\left[V^\dagger {\partial \phi \over \partial
H^\dagger}(H_0+Vt)\right]
\eqn\DefFtGt
$$
at $t=t_0$, then $H=H_0+Vt_0$ will be the lowest point of $\phi (H)$
in this cross section.
\item{iii)}
Starting from $t=0$, the Newton method applied to this function
$g(t)$ gives at the first iteration step
$$
t_1=-{g(0) \over g'(0)} \ ,
\ee
$$
as a nearer point to the zero $t_0$ of $g(t)$. We do not continue
this  Newton's iteration any further since even if $t_0$ is found
more exactly the point $H=H_0+Vt_0$ is merely the lowest point of
$\phi (H)$ inside this cross section. So we adopt $H_1=H_0+Vt_1$ as
a nearer point to a true stationary point of $\phi (H)$.
\item{iv)}
If $t_1$ is already small enough we consider $H_0$ is a stationary
point. Otherwise we take $H_1 \equiv H_0 + Vt_1$ as $H_0$ in the
step i) and repeat the procedure.

What we get by this iteration procedure is, logically speaking,
not a minimum point but a stationary point. But, in practice in this
calculation,
we actually obtained a minimum although it may not be a global minimum.

\section{Result}

We have run the above procedure for searching the stationary point
for the potential with various sets of parameters,
$M_{351}^2/M_{27}^2$ and $\Lambda ^2/M_{27}^2$; more explicitly,
we have swept the region
$0\leq M_{351}^2/M_{27}^2\leq 10^3$ and $40\leq \Lambda ^2/M_{27}^2\leq
10^3$.
(Note that $\Lambda ^2/M_{27}^2=\pi ^2$ is the critical value for the
symmetry breaking in the pure \twoseven\ interaction case.)
We have stored in total
about $10^4$ data of the stationary points $H_{\rm st}$
for the potentials with the parameters in this region, in particular,
in the region
$10^{-3}\leq M_{351}^2/M_{27}^2\leq 300$ and $\Lambda ^2/M_{27}^2=40,\
100$
in detail.

Using the obtained stationary point data $H_{\rm st}$, we have
calculated fermion masses and gauge boson masses on those vacua.
Fermion masses are calculated as eigenvalues of the squared mass
matrix $H_{\rm st}^\dagger H_{\rm st}$
and those of gauge bosons are as eigenvalues of the squared mass
matrix
$G = (G_{ab}) \equiv
(\tr[ T_aH_{\rm st}^\dagger +H_{\rm st}^\dagger T_a^{\rm T}])
(\tr[ T_b^*H_{\rm st}+H_{\rm st}T_b^\dagger ])$
where $T_a$'s are $E_6$ generators in the \twoseven\ representation,
whose explicit form is given in Appendix B.
We can judge the symmetry breaking pattern from those mass spectra
for each case.

The result of our numerical calculation is summarized as follows.
\item{1)}
When $M_{27}\le M_{351}$, namely, \twoseven\ 4-fermi interaction is
dominant, we found in every case of our search the following.
26 fermions has a degenerate mass and the rest
one fermion has another mass.
On the other hand,
52 gauge bosons are massless and the rest 26 has a degenerate non-zero
mass.
All these clearly imply that the symmetry breaking pattern
in this coupling region is
$$
\Esix\ \ \longrightarrow\ \ \Ffour \ .
\ee
$$
This completely agrees with the result obtained in the previous section,
despite that the latter was based on a bit non-rigorous arguments.
\item{2)}
When $M_{27}\ge M_{351}$, namely, \threefive\ 4-fermi interaction is
dominant, we found in every case the following.
All of the 27 fermions has a degenerate mass
while the gauge bosons become all massive but not degenerate at all.
The degenerate fermion spectrum implies that the symmetry breaking
pattern in this case is
$$
\Esix\ \ \longrightarrow\ \
Sp(8)\ \hbox{or}\ G_{\rm 2}\ \hbox{or}\ SU(3) \ ,
\ee
$$
agreeing again with the result of the previous analysis.
But it seems strange why all the gauge bosons are massive and
non-degenerate. If there remains some symmetry, the corresponding
gauge bosons should remain massless and the spectrum should show
some multiplet structure.
The reason why this strange thing happens is in the particular
nature in this breaking; namely, in this case, the three different
vacua with symmetries $Sp(8),\ G_{\rm 2}$ and $SU(3)$
are degenerate.  They place at different points in the potential
but realizes the same stationary value. Then, if there is a path
connecting these three points through which the potential is flat
(or almost flat within the calculation error), all the points on the
path realize the same stationary values but has no symmetries at all.
Nevertheless, the fermion mass degeneracy is still realized since
the present effective potential counts only the fermion vacuum energy
and the degeneracy of the potential value along the path means
the fermion mass degeneracy.
All the stationary points we found are such points on the path.
This is our interpretation, but we confirmed this by examining the
potential values realized by our stationary points. They all coincided
with $\phi _{Sp}=27f(v_0)$ which we obtained analytically in the
previous section by using Michel's conjecture.
\midinsert
\medskip
  \begintable
                 | $M_{27}\leq M_{351}$       | $M_{27}\geq  M_{351}$
\crthick
fermion mass     | \one+{\bf 26}           |  \twoseven         \cr
\ gauge bosson mass\ \ |\ {\bf52}(massless)+{\bf26}(massive)\ \
|\ \one(massive)$\times $78\ \
\endtable
\smallskip
\Table{2}{. \  Mass spectra found numerically for the cases
$M_{27}\leq M_{351}$ and $M_{27}\leq M_{351}$.}
\medskip\endinsert

\chapter{Summary and Conclusion}

We have analyzed an $E_6$ GUT model of a single generation of fermions
with strong 4-fermi interactions. The $E_6$ symmetry is found to be
broken spontaneously either to $F_4$ or to
$Sp(8)\ \hbox{or}\ G_{\rm 2}\ \hbox{or}\ SU(3)$ depending on which
of the 4-fermi coupling constants $G_{27}$ and $G_{351}$ in the
\twoseven /\threefive\ channels is stronger than the other.

In these symmetry breakings, the fermions turn to belong to {\it real}
representations of the residual symmetry and all of them get acquire
non-vanishing masses. Since
these masses are necessarily of the order of
the GUT symmetry breaking ($\sim 10^{16-17}$GeV), the present model
as it stands, unfortunately, turns out to be unrealistic as a GUT
model. The quarks and leptons belong to a chiral representation
of the standard gauge group and should remain massless at the GUT
scale.

We can easily understand the reason why all the fermions get
non-vanishing masses in the present model. As mentioned before,
our effective potential counts only the fermion one-loop vacuum energy.
But the fermion vacuum energy essentially comes from the energy
of Dirac's negative energy sea and hence is negative. So,
the more massive the fermions become, the more the vacuum energy
is lowered. Therefore the desirable symmetry breaking patterns,
such as down to $SU(3)\times SU(2)\times U(1)$ under which the fermions
are
chiral and remain massless, are necessarily disfavorable energetically.

This indicates that the vacuum energy coming from {\it bosons} should
play a central role in order for the present model to produce desirable
symmetry breaking patterns. Indeed, Harvey[\RHarvey] once considered
the
$E_6$ symmetry breaking in a Coleman-Weinberg like
spontaneous symmetry breaking scenario and found that $E_6$ is
broken down to $SO(10)$. There the main part of the potential in fact
came from the gauge boson loop contribution.

Alternatively, there may be another possibility
if we change the fermion content of the model. For instance[\RGeorgi],
we can
regard the three generations of quarks/leptons as merely survivals
from GUT world where $n+3$ generations and $n$ anti-generations
of fermions exist. Then,
when the dynamical GUT symmetry breaking occurs,
a variety of mixing can generally occur among those fermions, and
$n$ generations of fermions as a net number can get
acquire $O(M_{\rm GUT})$ masses leaving the usual quarks and leptons
massless. Since there are fermions which aquire the masses in this case,
 there is a possiblity that small contributions of the gauge boson loop
may be sufficient to realize such desirable breaking down to
chiral type symmetry.  This type of scenario is very interesting
also from the viewpoint of the origin of
Cabibbo-Kobayashi-Maskawa mixing as well as of the stability of proton.

\acknowledge
The authors would like to thank Profs. T.~Maskawa and M.~Bando
for their valuable comments and discussions.
T.K. is supported in part by the Grant-in-Aid for Scientific Research
(\#04640292) from the Ministtry of Education, Science and Culture.

\APPENDIX{A}{A.\ $SO(10)$  $\gamma $-Matrices}

For any $SO(2n)$, the $\gamma $-matrices
$\gam{2n}_M\ (M = 1, 2, \cdots , 2n)$
satisfying $\gam{2n}_M\gam{2n}_N+\gam{2n}_N\gam{2n}_M = 2\delta _{MN}$
take the form
$$
\gam{2n}_M = \mymatrix{0}{\big(\sgm{2n}_M\big)_{\alpha \beta }}
{\big(\sgm{2n}_M^\dagger\big)^{\alpha \beta }}{0}
\qquad {\rm on} \quad \myvector{\xi _\beta }{\eta ^\beta }\ ,
\ee
$$
where $\xi _\alpha $ and $\eta ^\alpha $ are $2^{{n\over
2}-1}$-component Weyl spinors
with chiralities $\gam{2n}_{2n+1}=+1$ and $-1$, respectively.
The superscript on the left shoulder indicates the dimension $2n$
of $SO(2n)$.
$\sigma $'s are the $\gamma $-matrices in the Weyl spinor basis.

Totally anti-symmetric multi-indexed $\gamma $-matrices
$\gam{2n}_{M_1M_2\cdots M_k}$ are defined by
$$
\eqalign{
\gam{2n}_{M_1M_2\cdots M_k} &= {1\over k!}\left( \gam{2n}_{M_1}\!
\gam{2n}_{M_2}\cdots \gam{2n}_{M_k} + \hbox{(anti-symmetrization)}
\right) \cr
&\equiv  \cases{
\mymatrix{\sgm{2n}_{M_1M_2\cdots M_k}}{0}{0}
{\barsgm{2n}_{M_1M_2\cdots M_k}}
 & for $k=$ even \cr
\mymatrix{0}{\sgm{2n}_{M_1M_2\cdots M_k}}
{\barsgm{2n}_{M_1M_2\cdots M_k}}{0}
 & for $k=$ odd \cr } \cr
\sgm{2n}_{M_1M_2\cdots M_k} &= {1\over k!}\left(
\sgm{2n}_{M_1}\sgm{2n}^\dagger_{M_2}\sgm{2n}_{M_3}\sgm{2n}^\dagger_{M_4}
\cdots\sgm{2n}^{(\dagger)}_{M_k} + \hbox{(anti-symmetrization)}\right)
\ , \cr
\barsgm{2n}_{M_1M_2\cdots M_k} &= {1\over k!}\left(
\sgm{2n}^\dagger_{M_1}\sgm{2n}_{M_2}\sgm{2n}^\dagger_{M_3}\sgm{2n}_{M4}
\cdots\sgm{2n}^{(\dagger)}_{M_k} + \hbox{(anti-symmetrization)}\right)
\ . \cr
}\ee
$$
The $SO(2n)$ generators $T_{MN}$ satisfying
$[T_{MN},\ T_{KL}] = -i\big( \delta _{NK}T_{ML} +
$ (anti-symmetrization)$\big)$ are expressed in this spinor
representation by the matrix
$$
\Sgm{2n}_{MN}\equiv {1\over 2i}\gam{2n}_{MN}
= {1\over 2i}\mymatrix{\sgm{2n}_{MN}}{0}{0}{\barsgm{2n}_{MN}} \ .
\eqn\eqSOGENERATOR
$$
The charge conjugation matrix $\CC{2n}$ exists such that\refmark{\refKT}
$$
\eqalign{
&\CC{2n}\gam{2n}_M \CC{2n}^{-1} = \eta  \gam{2n}_M^{\rm T} \cr
& \CC{2n}^\dagger\CC{2n} = 1\ ,
\quad  \CC{2n}^{\rm T} = \epsilon  \CC{2n} \quad
{\rm with}\ \ \epsilon =\cos{n\pi \over 2}+\eta \sin{n\pi \over 2} \cr
}\eqn\eqCC
$$
for either choice of $\eta =\pm 1$, where the superscript T denotes
transposed. Henceforth we always choose $\eta =+1$ for convenience.

\section{$SO(6)$}
We first construct $SO(6)$ $\gamma $-matrices in the Weyl spinor basis:
it is convenient to take the $4\times 4$\  $\sgm{6}_m $ matrices as
$$
\eqalign{
&\sgm{6}_m = \left(\  \sgm{6}_{i=1,2,3},
\ \sgm{6}_{i+3=4,5,6}\  \right)
\cr
& \cases{
\big( \sgm{6}_i \big)_{\alpha \beta } =  \varepsilon _{i4\alpha \beta }
+ \delta _{\alpha \beta }^{i4} \cr
\big( \sgm{6}_{i+3} \big)_{\alpha \beta } =  i\left(
\varepsilon _{i4\alpha \beta } - \delta _{\alpha \beta }^{i4} \right)
\cr}
\ ,\cr
}\ee
$$
where $\varepsilon _{\alpha \beta \gamma \delta }$ is rank-4 totally
anti-symmetric tensor and
$\delta _{\alpha \beta }^{\gamma \delta }$ is multi-index
anti-symmetric Kronecker's
delta defined by
$\delta _{\alpha \beta }^{\gamma \delta }\equiv \delta _\alpha ^\gamma
\delta _\beta ^\delta -\delta _\alpha ^\delta \delta _\beta ^\gamma $.
The index $i$ here, running over 1, 2, 3,  will correspond to the
color index of $SU(3)\subset SO(6)$.

These $\sgm{6}_m\ (m=1,2, \cdots ,6)$ possess the following properties:
$$
\eqalign{
&\sgm{6}_m = - \sgm{6}_m^{\rm T} \qquad ( \hbox{anti-symmetric}) \cr
&\big(\sgm{6}_m\big)_{\alpha \beta } = -{1\over 2}\varepsilon _{\alpha
\beta \gamma \delta }
\big( \sgm{6}^\dagger_m \big)^{\gamma \delta }
\qquad (\hbox{anti-selfduality}) \cr
&{1\over 2}\big(\sgm{6}_m\big)_{\alpha \beta
}\big(\sgm{6}_m^\dagger\big)^{\gamma \delta }
= - \delta _{\alpha \beta }^{\gamma \delta } \ \ \leftrightarrow \ \
{1\over 4} \tr \big( \sgm{6}_m\sgm{6}_n^\dagger\big)=\delta _{mn} \cr
&{1\over 2}\big(\sgm{6}_m\big)_{\alpha \beta
}\big(\sgm{6}_m\big)_{\gamma \delta }
= \varepsilon _{\alpha \beta \gamma \delta }\ . \cr
}\ee
$$
An $SO(6)$-vector $V_m$ is eqiuvalent to a rank-2 antisymmetric tensor
$V_{[\alpha \beta ]}$ of $SU(4)$; they are related with each other via
$$
V_{[\alpha \beta ]}= {1\over \sqrt2}(\sgm{6}_m)_{\alpha \beta } V_m
 \ \leftrightarrow \
V_m
= {1\over 2}\Big({1\over \sqrt2}\sgm{6}^\dagger_m\Big)^{\alpha \beta }
V_{[\alpha \beta ]}
\ .
\ee
$$
Decomposition of the $SO(6)$ vector $V_m$ into ${\bf 3}+{\bf 3}^*$
under the color group $SU(3) \subset SU(4)\simeq SO(6)$ is given by
$$
\eqalign{
&{\bf 3}: \quad V_{[i4]} = {1\over \sqrt2}\big(V_i-iV_{i+3}\big) \ .\cr
&{\bf 3}^*: \quad {1\over 2}\varepsilon ^{ijk}V_{[jk]}
= {1\over \sqrt2}\big(V_i+iV_{i+3}\big) \cr
}\ee
$$
The $SO(6)$ generators are given by the general expression
\eqSOGENERATOR, which defines $\sgm{6}_{mn}$ and $\barsgm{6}_{mn}$.
Then the 15 matrices $\sgm{6}_{mn}$ ($m,n=1,\cdots ,6$) together with
a unit matrix span a complete set of $4\times 4$ matrices and satisfy
the following completeness relation:
$$
{1\over 4}(1)_\alpha ^{\ \gamma }(1)_\beta ^{\ \delta }
+ {1\over 2}(\sgm{6}_{mn})_\alpha ^{\ \gamma }(\sgm{6}_{mn})_\beta ^{\
\delta }
= \delta _\alpha ^\delta \,\delta _\beta ^\gamma  \ .
\ee
$$
The charge conjugation matrix $\CC{6}$ defined generally
in \eqCC\ is now given by
$$
\CC{6}=\mymatrix{0}{-\uni{4}}{\uni{4}}{0} = -\CC{6}^{\rm T} \ ,
\ee
$$
with $\uni{m}$ denoting $m\times m$ unit matrix.

\section{$SO(4)$}
$SO(4)$ $\gamma $-matrices $\sgm{4}_\mu \ (\mu =7,8,9,0)$
in the Weyl spinor basis are $2\times 2$ matrices
which we take as follows:
$$
\sgm{4}_\mu =\big( -i\sigma _1,  -i\sigma _2,  -i\sigma _3, \uni{2}
\big)
\ee
$$
with $\sigma _{1,2,3}$ being the Pauli matrices. Then $SO(4)$ generators
$\Sgm{4}_{\mu \nu }$ defined by \eqSOGENERATOR\ split into
$3+3$ generators of $SU(2)_{\rm L}\times SU(2)_{\rm R}\simeq SO(4)$:
$$
\eqalign{
&\Sigma _{{\rm L}i} = {1\over 2}\left( {1\over 2}\varepsilon
_{ijk}\Sgm{4}_{j+6,k+6}
+\Sgm{4}_{0,i+6} \right) = \mymatrix{{1\over 2}\sigma _k}{}{}{0} \cr
&\Sigma _{{\rm R}i} = {1\over 2}\left( {1\over 2}\varepsilon
_{ijk}\Sgm{4}_{j+6,k+6}
-\Sgm{4}_{0,i+6} \right) = \mymatrix{0}{}{}{{1\over 2}\sigma _k} \ .\cr
}\ee
$$
The charge conjugation matrix $\CC{4}$ is given by
$$
\CC{4}=\mymatrix{i\sigma _2}{0}{0}{i\sigma _2} = - \CC{4}^{\rm T} \ .
\ee
$$

\section{$SO(10)$}
$SO(10)$ $\gamma $-matrices $\gam{10}_M\ (M=1,2,\cdots ,9,0)$ are
constructed
by a tensor product of the $SO(6)$ and $SO(4)$ $\gamma $-matrices as
follows:
$$
\gam{10}_M = \cases{
\gam{6}_m \otimes \gam{4}_5 & for $M=m=1,2,\cdots, 6$ \cr
\ \ \uni{4} \otimes \gam{4}_\mu  & for $M=\mu=7,8,9,0$ \cr} \ ,
\ \ {\rm with\ } \gam{4}_5 = \mymatrix{\uni{2}}{0}{0}{-\uni{2}}\ .
\ee
$$
Then the $\gamma $-matrices in the Weyl basis, $\sigma _M$, for which
we omit the
superscript 10 implying $SO(10)$ for notational simplicity,  are
$8\times 8$ matrices taking the following form:
$$
\sigma _{M=m} = \mymatrix{0}{\sgm{6}_m\otimes\uni{2}}
{-\sgm{6}^\dagger_m\otimes\uni{2}}{0} \ ,
\qquad
\sigma _{M=\mu } = \mymatrix{\uni{4}\otimes\sgm{4}_\mu }{0}
{0}{\uni{4}\otimes\sgm{4}^\dagger_\mu } \ .
\ee
$$
The charge conjugation matrix $\CC{10}$ takes the form
$$
\CC{10}=\CC{6}\otimes\CC{4} = \mymatrix{0}{C}{C}{0} \ ,
\ee
$$
where $C$ is $8\times 8$ matrix given by
$$
C=\mymatrix{0}{-\uni{4}\otimes i\sigma _2}
{\uni{4}\otimes i\sigma _2}{0} = C^{\rm T} = C^{-1} = C^\dagger \ .
\ee
$$

{}From $\CC{10}^{\rm T} = \CC{10}$, it follows that
the matrices $\CC{10}\gam{10}_M$ are symmetric, and so are
$C\sigma _M^{(\dagger)}$ and $\sigma _M^{(\dagger)}C$.
Similarly we see that $C\bar\sigma _{M_1M_2\cdots M_5}$ and
$\sigma _{M_1M_2\cdots M_5}C$ are symmetric. Since they are selfdual,
$$
C\bar\sigma _{M_1\cdots M_5}={1\over 5!}i
\varepsilon _{M_1\cdots M_5N_1\cdots N_5}C\bar\sigma _{N_1\cdots N_5},
\ee
$$
they give ${}_{10}C_5/2=126$ symmetric matrices, and hence, together
with the ten $C\sigma ^\dagger_M$ matrices, span a complete set in the
space
of $16\times 16$ symmetric matrices; the completeness relation reads
$$
2^{-4}\left[ (\sigma _MC)_{\alpha \beta }(C\sigma ^\dagger_M)^{\gamma
\delta } +
 {1\over 2\cdot 5!}(\sigma _{M_1\cdots M_5}C)_{\alpha \beta }
(C\bar\sigma _{M_1\cdots M_5})^{\gamma \delta }
\right] = {1\over 2}\big(\delta _\alpha ^\delta \delta _\beta ^\gamma
+\delta _\alpha ^\gamma \delta _\beta ^\delta  \big) \ ,
\ee
$$
because of the normalization condition
$$
\eqalign{
&2^{-4}\tr\big(C\sigma ^\dagger_M\sigma _NC\big) = \delta _{MN} \cr
&2^{-4}\tr\big(C\bar\sigma _{M_1\cdots M_5}\sigma _{N_1\cdots
N_5}C\big)
= \delta _{M_1\cdots M_5}^{N_1\cdots N_5} + i\varepsilon _{M_1\cdots
M_5N_1\cdots N_5} \ .
\cr
}\ee
$$
In the same way  ${}_{10}C_3=120$ matrices $C\bar\sigma _{M_1M_2M_3}$
(or, $\sigma _{M_1M_2M_3}C$) turn out to give a complete set of
$16\times 16$
anti-symmetric matrices so that
$$
2^{-4}{1\over 3!}(\sigma _{M_1M_2M_3}C)_{\alpha \beta }
(C\bar\sigma _{M_1M_2M_3})^{\gamma \delta }
 = {1\over 2}\big(\delta _\alpha ^\delta \delta _\beta ^\gamma -\delta
_\alpha ^\gamma \delta _\beta ^\delta  \big) \ .
\ee
$$

\APPENDIX{B}{B.\ Some Representations of $E_6$ }

The $E_6$ algebra is most easily expressed by refering to its
maximal subgroup $SO(10)\times U(1)$. The generators are given by
16 $SO(10)$ Weyl-spinor generators $E_\alpha $ $(\alpha =1,\cdots ,16)$
and
their complex conugates $\bar E^\alpha =(E_\alpha )^\dagger$ in
addition to the
45 $SO(10)$ generators $T_{MN}$ and one $U(1)$ generator $T$.
The algebra is given by\refmark{\refIKK}
$$
\eqalign{
[ \,T_{MN}, \,T_{KL}\, ] &= -i\big( \delta _{NK}T_{ML}
+\delta _{ML}T_{NK} -\delta _{MK}T_{NL} -\delta _{NL}T_{MK} \big) \ ,
\cr
[ \,T_{MN}, \,\myvector{E_\alpha }{\bar E^\alpha }\, ]
&= - \mymatrix{(\sigma _{MN})_\alpha ^{\ \beta }}{0}{0}{(-\sigma
_{MN}^*)^\alpha _{\ \beta }}
\myvector{E_\beta }{\bar E^\beta } \ , \cr
[ \,T, \,\myvector{E_\alpha }{\bar E^\alpha }\, ]
&= {\sqrt3\over 2}
\myvector{E_\alpha }{-\bar E^\alpha } \ , \cr
[ \,E_\alpha , \,\bar E^\beta \, ] &= -{1\over 2}(\sigma _{MN})_\alpha
^{\ \beta }T_{MN}
+ {\sqrt3\over 2}\delta _\alpha ^\beta T \ . \cr
}\eqn\eqALGEBRA
$$

The simplest representation of $E_6$ is {\bf 27} which is decomposed
into ${\bf 1}_4  + {\bf 16}_{1} +{\bf 10}_{-2}$ under the maximal
subgroup $SO(10)\times U(1)$. (The suffices denote the value of $U(1)$
charge $2\sqrt3 T$.) So the {\bf 27} representation can be denoted as
$\psi _A\equiv  (\psi _0, \psi _\alpha , \psi _M)$ with $\alpha $ and
$M$ being $SO(10)$
(Weyl-)spinor and vector indices, respectively.
The $E_6$ generators act on this representation as\refmark{\refIKK}
$$
\eqalign{
\big(\theta T+&{1\over 2}\theta _{KL}T_{KL}+\bar\epsilon ^\gamma
E_\gamma +\bar E^\gamma \epsilon _\gamma \big)
\sanvector{\psi _0}{\psi _\alpha }{\psi _M} \cr
&= \sanmatrix{{2\over \sqrt3}\theta }{\bar\epsilon ^\beta }{0}{\epsilon
_\alpha }
{{1\over 2}\theta _{KL}(\sigma _{KL})_\alpha ^{\ \beta }+{1\over
2\sqrt3}\theta \delta _\alpha ^\beta }
{-{1\over \sqrt2}(\bar\epsilon \sigma _NC)_\alpha }{0}{-{1\over
\sqrt2}(C\sigma ^\dagger_M\epsilon )^\beta }
{-i\theta _{MN}-{1\over \sqrt3}\theta \delta _{MN}}
\sanvector{\psi _0}{\psi _\beta }{\psi _N} \cr
}\eqn\eqTWOSEVENREPR
$$
To check that this representation for the $E_6$
generators really satisfies the algebra \eqALGEBRA, we need the
following identities for the $SO(10)$ $\gamma $-matrices:
$$
\eqalign{
&C\bar\sigma _{MN}C = -\sigma ^{\rm T}_{AB}\ , \qquad
C\bar\sigma _{M_1M_2M_3M_4}C = \sigma ^{\rm T}_{M_1M_2M_3M_4}\ , \cr
&\sigma _{MN}\sigma _K-\sigma _K\bar\sigma _{MN}=i\big(\delta
_{MK}\sigma _N-\delta _{NK}\sigma _M\big) \ , \cr
& {1\over 2}(C\sigma _M^\dagger)^{\beta \gamma }(\sigma _MC)_{\delta
\alpha }-\delta _\alpha ^\gamma \delta _\delta ^\beta  =
{1\over 4}\delta ^{\ \beta }_\alpha \delta ^{\ \gamma }_\delta
-{1\over 2}(\sigma _{MN})_\alpha ^{\ \beta }(\sigma _{MN})_\delta ^{\
\gamma } \ . \cr
}\ee
$$
The last identity follows from the Fierz transformation of the LHS:
$$
\eqalign{
&\delta _\alpha ^\gamma \delta _\delta ^\beta  = 2^{-4}\left[ (1)^{\
\beta }_\alpha (1)^{\ \gamma }_\delta
-{1\over 2}(2i\sigma _{MN})^{\ \beta }_\alpha (2i\sigma _{MN})^{\
\gamma }_\delta
+{1\over 4!}(\sigma _{M_1\cdots M_4})^{\ \beta }_\alpha (\sigma
_{M_1\cdots M_4})^{\ \gamma }_\delta
\right] \ , \cr
& (C\sigma _M^\dagger)^{\beta \gamma }(\sigma _MC)_{\delta \alpha } =
2^{-4} \bigg[
(C1C)^\beta _{\ \alpha }(\sigma _K1\sigma ^\dagger_K)^{\ \gamma
}_\delta  \cr
&\quad
-{1\over 2}(C2i\bar\sigma _{MN}C)^\beta _{\ \alpha }
(\sigma _K2i\bar\sigma _{MN}\sigma _K^\dagger)^{\ \gamma }_\delta
+{1\over 4!}(C\bar\sigma _{M_1\cdots M_4}C)^\beta _{\ \alpha }
(\sigma _K\bar\sigma _{M_1\cdots M_4}\sigma ^\dagger_K)^{\ \gamma
}_\delta
\bigg] \ , \cr
&\sigma _K\bar\sigma _{MN}\sigma _K^\dagger=6\sigma _{MN}\ , \qquad
\sigma _K\bar\sigma _{M_1\cdots M_4}\sigma _K^\dagger=2\sigma
_{M_1\cdots M_4}\ . \cr
}\ee
$$

Tensor product of two {\bf 27} representations gives
$$
{\bf 27}\times {\bf 27} = \bar{\bf 27}_{\rm S} + \bar{\bf 351}_{\rm S}
+ \bar{\bf 351}'_{\rm A} \ .
\ee
$$
This implies that there is an invariant tensor $\Gamma ^{ABC}$ which
gives $\bar{\bf 27}$ from ${\bf 27}\times {\bf 27}$:
$$
\bar\Psi ^A = \Gamma ^{ABC}\psi _B\psi _C \ .
\ee
$$
This $\Gamma ^{ABC}$ is found to be given by
$$
\Gamma ^{ABC}  :  \ \hbox{totally symmetric}
\quad \   \cases{ \Gamma ^{0MN} = \delta _{MN}  \cr
\Gamma ^{M\alpha \beta }= {1\over \sqrt2}(C\sigma _M^\dagger)^{\alpha
\beta } \cr
\hbox{otherwise} \ \  0 \ \ ,}
\ee
$$
or equivalently, in terms of the components of $\bar\Psi ^A$,
$$
\eqalign{
&\bar\Psi ^0 = \psi _M\psi _M \cr
&\bar\Psi ^M = {1\over \sqrt2}\psi ^{\rm T}C\sigma _M^\dagger\psi
+2\psi _0\psi _M \cr
&\bar\Psi ^\alpha = \sqrt2 \,\psi _M (C\sigma _M^\dagger\psi )^\alpha
\ .\cr
}\ee
$$
To check that this $\bar\Psi $ transforms correctly as $\bar{\bf 27}$,
we need an identity:
$$
(\epsilon ^{\rm T}C\sigma _M^\dagger\psi )(\psi ^{\rm T}C\sigma
_M^\dagger\eta )
= -{1\over 2}(\epsilon ^{\rm T}C\sigma _M\eta )(\psi ^{\rm T}C\sigma
_M^\dagger\psi ) \ ,
\ee
$$
which follows from Fierzing $\sigma _M^\dagger\psi $ and $\eta $ and
using
$\sigma _M^\dagger\sigma _K \sigma _M^\dagger = -8 \sigma _K^\dagger$
and
$\sigma _M^\dagger\sigma _{K_1\cdots K_5} \sigma _M^\dagger = 0$.

The {\bf 351} can be represented by a symmetric tensor $\Phi ^{}$ with
two $\bar{\bf 27}$ indices $A$ and $B$:
$$
\Phi ^{AB} = \sanmatrix
{\Phi ^0({\bf 1})} {{1\over \sqrt2}\Phi ^\beta (\bar{\bf16})}
                        {{1\over \sqrt2}\Phi ^N({\bf10})}
{{1\over \sqrt2}\Phi ^\alpha (\bar{\bf16})}  {\Phi ^{\alpha \beta
}(\bar{\bf126})}
                        {{1\over \sqrt2}\Phi ^{\alpha N}({\bf144})}
{{1\over \sqrt2}\Phi ^M({\bf10})} {{1\over \sqrt2}\Phi ^{M\beta }({\bf
144})}
                        {\Phi ^{MN}({\bf54})} \ ,
\eqn\eqSANGOITI
$$
where the argument in each entry denotes the dimension under $SO(10)$.
The previous {\bf 27} representation $\psi ^A$ can also be imbedded
into
a symmetric matrix using the invariant symmetric tensor $\Gamma ^{ABC}$:
$$
\Gamma ^{ABC}\psi _C =
\sanmatrix{0}{0}{\psi _N}
{0}{{1\over \sqrt2}\psi _K(C\sigma _K^\dagger)^{\alpha \beta }}
           {{1\over \sqrt2}(C\sigma _N^\dagger\psi )^\alpha }
{\psi_M}{{1\over \sqrt2}(C\sigma _M^\dagger\psi )^\beta }{\delta
_{MN}\psi _0} \ .
\eqn\eqTWOSEVEN
$$
Since symmetric tensor product of two $\bar{\bf 27}$ is
either {\bf 27} or {\bf 351}, the {\bf 351} matrix $\Phi ^{AB}$ can be
characterized as a general symmetric matrix which contains no {\bf 27}
components of the form \eqTWOSEVEN: therefore, the component
$\Phi ^{MN}$ should be traceless, $\Phi ^{MM}=0$;
$\Phi ^{\alpha \beta }$ should contain no
$SO(10)$ vector components,
$(\sigma _MC)_{\alpha \beta }\Phi ^{\alpha \beta }=0$;
$\Phi ^{M\beta }$ should be $\gamma $-traceless, $(\sigma _MC)_{\alpha
\beta }\Phi ^{M\beta }=0$.
But, as a matter of course,
these conditions are nothing but the requirements
that those entries be irreducible
representations under $SO(10)$ as indicated in the arguments in
\eqSANGOITI.

\APPENDIX{\HurokuMaxLittle }{\HurokuMaxLittle.\ Maximal Little Group}

A little group of a representation vector $\phi$ of a group $G$
is defined by
$$
H_\phi \equiv \Big\{\ g\ \Big|\ g\phi=\phi, g\in G \Big\} \ .
\ee
$$
This little group depends not only on the representation but also
on the vector $\phi$ itself.

Consider a single irreducible representation $R$ or a self-conjugate
pair $R+R^*$ of a complex irredicible representation $R$.
For this representation $R$, many little groups appear as
the vector $\phi $ varies in the representation $R$ with the length
$|\phi |(\not=0)$ kept fixed.
A little group $H$ is called {\it maximal} if there is no $\phi $ with
little group $H_\phi $ satisfying $G\supset H_\phi \supset H$.

Some examples of $E_6$ maximal little groups are given in the
following Table.
\midinsert
\medskip
  \begintable
   $R$      | Maximal little groups  \crthick
 {\bf 78}   | \ $SU(6)\times U(1), \ SO(10)\times U(1), \
SU(5)\times SU(2)\times U(1), \ [SU(3)]^2\times SU(2)\times U(1)$ \  \cr
 {\bf 27}   | $SO(10), \ F_4$   \cr
 \ {\bf 351}\  | $SO(10), \ F_4,\  Sp(8),\ G_2,\ SU(3),\
SU(4)\times SU(2)$
\endtable
\smallskip
\Table{3}{. \  Maximal little groups for the representations
$R\ (+R^*)$.}
\medskip\endinsert

\endpage

\refout

\bye